\documentclass{ECOC-author}

\usepackage{tikz}
\usepackage{pgfplots}
\usetikzlibrary{colorbrewer}
\usetikzlibrary{arrows}
\usetikzlibrary{shapes,decorations}
\usepackage{caption}
\usepackage{amsmath,amssymb}
\usepackage{mathtools, cuted}
\usepackage{graphicx}
\newcommand{\appropto}{\mathrel{\vcenter{
  \offinterlineskip\halign{\hfil$##$\cr
    \propto\cr\noalign{\kern2pt}\sim\cr\noalign{\kern-2pt}}}}}

\begin{document}

\title{On the Relationship Between Network Topology and Throughput in Mesh Optical Networks}

\author{Daniel Semrau\ad{1}\corr, Shahzaib Durrani\ad{1}, Georgios Zervas\ad{1}, Robert I. Killey\ad{1}, Polina Bayvel\ad{1}}

\address{\add{1}{Optical Networks Group, Department of Electronic and Electrical Engineering, UCL (University College London), Torrington Place, London WC1E 7JE, United Kingdom}
\email{uceedfs@ucl.ac.uk}}


\begin{abstract}
The relationship between topology and network throughput of arbitrarily-connected mesh networks is studied. Taking into account nonlinear channel properties, it is shown that throughput decreases logarithmically with physical network size with minor dependence on network ellipticity.
\end{abstract}

\maketitle

\section{Introduction}
Optical networks underpin the infrastructure of the modern Internet. To maximise the throughput of these networks, physical propagation effects must be taken into account. Thus, the network throughput becomes a nonlinear function of fibre parameters, link distances and lightpath configurations, and networks can no longer be only characterised only by logical topologies, i.e. a set of vertices and edges. 
\par 
\ 
The development of low complexity transmission models of nonlinear fibre propagation enabled their incorporation in network algorithms. As a consequence, various studies have suggested new optimisation strategies for wavelength and route assignments \cite{nw1,nw2, nw4, nw7}, launch power optimisations and the use of various modulation formats and code rates \cite{Ives,nw3} within the same network, tailored to the lightpath quality. 
\par
\ 
While many works analyse specific topologies, often derived from existing/published networks, more general relationships between \textit{physical} network topology and throughput are largely nonexistant. Most work \cite{Baroni,nw5} analysed network performance as a function of node, link number and connectivity matrix in the absence of physical transmission properties. However, physical transmission properties must now be included in order to calculate both throughput as well as derive scaling laws that lead to a more intelligent use of current optical networks.
\par 
\ 
In this work, arbitrarily-connected mesh optical networks (ACMN) were analysed to analyse the relationship between physical topology properties and network throughput. The focus of the work was two network properties - scaling the network size and paramterising network ellipticity. 
\begin{figure}[t]
\centering

\begin{tikzpicture}



\node (topology_generation_group) [rectangle,minimum height=4.7cm, xshift=1.5 cm, yshift=.1cm,minimum width=5.1cm,draw,dashed,thick,align=left] {};

\node[anchor=north west,inner sep=0pt, yshift=-0.1cm, xshift=0.1cm] at (topology_generation_group.north west) 
    {\footnotesize Topology generation:};

\node (distance_pdf) [rectangle,minimum height=2.5cm, minimum width=3.2cm,thick, yshift=1.35cm,xshift=-1.7cm] at (topology_generation_group.south east) {};

\node (generate_1000_RCN) [rectangle,draw,thick, align=center, xshift=-2.4cm] at (distance_pdf)  {\footnotesize Generate \\ \footnotesize ACMNs};

\node (assign_link_distances) [rectangle,draw,thick, align=center, xshift=-2.4cm, yshift=2.5 cm] at (distance_pdf)  {\footnotesize Assign link\\ \footnotesize distances};

\node (dataset1) [rectangle,minimum height=1.2cm, minimum width=1.6cm,draw,thick, xshift=3.1 cm, yshift=-.2 cm] at (assign_link_distances) {};
\node (dataset2) [rectangle,fill=white,minimum height=1.2cm, minimum width=1.6cm,draw,thick, xshift=3 cm, yshift=-.1 cm] at (assign_link_distances) {};
\node (dataset3) [rectangle,fill=white,minimum height=1.2cm, minimum width=1.6cm,draw,thick, xshift=2.9 cm] at (assign_link_distances) {};

\node () [rectangle,thick, xshift=-0.5cm,yshift=-0.15cm] at (dataset1.south east) {\scriptsize $\times 100000$};
\node () [rectangle,thick, xshift=0.35cm,yshift=-0.15cm] at (generate_1000_RCN.south west) {\scriptsize $\times 1000$};
\node () [rectangle,thick, xshift=0.3cm,yshift=-0.15cm] at (assign_link_distances.south west) {\scriptsize $\times 100$};


\node (top1) [circle,fill=black,outer sep=0.03cm,inner sep=0.03cm,draw,thick,xshift=-.55cm,yshift=.4cm] at (dataset3) {};
\node (top2) [circle,fill=black,outer sep=0.03cm,inner sep=0.03cm,draw,thick,xshift=-.45cm,yshift=0cm] at (dataset3) {};
\node (top3) [circle,fill=black,outer sep=0.03cm,inner sep=0.03cm,draw,thick,xshift=-.15cm,yshift=.3cm] at (dataset3) {};
\node (top4) [circle,fill=black,outer sep=0.03cm,inner sep=0.03cm,draw,thick,xshift=-.25cm,yshift=-.3cm] at (dataset3) {};
\node (top5) [circle,fill=black,outer sep=0.03cm,inner sep=0.03cm,draw,thick,xshift=.15cm,yshift=0cm] at (dataset3) {};
\node (top6)[circle,fill=black,outer sep=0.03cm,inner sep=0.03cm,draw,thick,xshift=.25cm,yshift=-.35cm] at (dataset3) {};
\node (top7) [circle,fill=black,outer sep=0.03cm,inner sep=0.03cm,draw,thick,xshift=.35cm,yshift=.35cm] at (dataset3) {};\node (top8) [circle,fill=black,outer sep=0.03cm,inner sep=0.03cm,draw,thick,xshift=.55cm,yshift=0cm] at (dataset3) {};

\draw[-,black] (top1.center) -- (top2.center);
\draw[-,black] (top2.center) -- (top4.center);
\draw[-,black] (top1.center) -- (top3.center);
\draw[-,black] (top3.center) -- (top7.center);
\draw[-,black] (top5.center) -- (top7.center);
\draw[-,black] (top5.center) -- (top3.center);
\draw[-,black] (top5.center) -- (top6.center);
\draw[-,black] (top4.center) -- (top6.center);
\draw[-,black] (top8.center) -- (top6.center);
\draw[-,black] (top8.center) -- (top7.center);

\node[anchor=north west,inner sep=0pt, yshift=-0.1cm, xshift=5.3cm, align=center] at (topology_generation_group.north west) 
    {\footnotesize For each topology:};

\node (for_each_topology1) [inner sep=0pt,outer sep=0pt,yshift=-0.1 cm, xshift=.1 cm] at (topology_generation_group.south east) {};
\node (for_each_topology2) [inner sep=0pt,outer sep=0pt,yshift=-0.1 cm] at (topology_generation_group.south west) {};
\node (for_each_topology3) [inner sep=0pt,outer sep=0pt,yshift=-2.8 cm] at (for_each_topology2) {};
\node (for_each_topology4) [inner sep=0pt,outer sep=0pt,xshift=8.5 cm] at (for_each_topology3) {};
\node (for_each_topology5) [inner sep=0pt,outer sep=0pt] at ([xshift=3.4cm]topology_generation_group.north east) {};

\draw[-,dashed,thick,black] ([xshift=.1cm]topology_generation_group.north east) -- (for_each_topology5.center);
\draw[-,dashed,thick,black] ([xshift=.1cm]topology_generation_group.north east) -- (for_each_topology1.center);
\draw[-,dashed,thick,black] (for_each_topology1.center) -- (for_each_topology2.center);
\draw[-,dashed,thick,black] (for_each_topology2.center) -- (for_each_topology3.center);
\draw[-,dashed,thick,black] (for_each_topology3.center) -- (for_each_topology4.center);
\draw[-,dashed,thick,black] (for_each_topology5.center) -- (for_each_topology4.center);

\node (transform) [rectangle,draw,thick, align=center,yshift=-1.2cm] at  ([xshift=1.75cm]topology_generation_group.north east) {\footnotesize Compute NSR weights\\ \footnotesize from link distances \\ \footnotesize using GN model};

\node (route_selection) [rectangle,draw,thick, align=center,yshift=-3.3cm] at  ([xshift=1.75cm]topology_generation_group.north east) {\footnotesize For each node-pair: \\ \footnotesize compute paths with \\ \footnotesize capacity \\ \footnotesize $C_i\geq x\cdot\underset{\forall i}{\max}\left(C_i\right)$ \\ \footnotesize using k-shortest paths};


\node (RWA) [rectangle,draw,thick, align=center,yshift=-6.2cm] at  ([xshift=1.75cm]topology_generation_group.north east) {\footnotesize Assign  $N_\lambda$ ligth paths to \\ \footnotesize each node-pair such that\\ \footnotesize congestion is minimised};


\node (if) [rectangle,draw,thick, align=center,xshift=-2.6 cm] at (RWA) {\footnotesize C-band \\ \footnotesize reached on \\ \footnotesize any link?};

\node (T) [rectangle,draw,thick, align=center,xshift=-5.5 cm] at (RWA) {\footnotesize Compute light path\\ \footnotesize capacities based \\ \footnotesize on RWA using \\ \footnotesize GN model};

\node (no_loop1) [inner sep=0pt,outer sep=0pt,yshift=-1.3 cm] at (if) {};
\node (no_loop2) [inner sep=0pt,outer sep=0pt,yshift=-1.3 cm] at (RWA) {};

\draw[->,thick,black] (generate_1000_RCN) -- (assign_link_distances);
\draw[->,thick,black] (assign_link_distances) -- (dataset3);
\draw[->,thick,black] (distance_pdf) -- (assign_link_distances);
\draw[->,thick,black] (transform) -- (route_selection);
\draw[->,thick,black] (RWA) -- (if);
\draw[->,thick,black] (route_selection) -- (RWA) node[align=center,draw=none,fill=white,font=\scriptsize,midway] {provides `equal-cost` paths \\ for each node-pair};
\draw[->,thick,black] (if) -- (T) node[draw=none,fill=white,font=\scriptsize,midway] {yes};
\draw[-,thick,black] (if) -- (no_loop1.center) node[draw=none,fill=white,font=\scriptsize,midway] {no};
\draw[-,thick,black] (no_loop1.center) -- (no_loop2.center);
\draw[->,thick,black] (no_loop2.center) -- node [right] {\footnotesize $N_\lambda$++} (RWA);
\begin{axis}[%
width=2.2cm,
height=1.7cm,
at={(1.7cm,-1.5cm)},
unbounded coords=jump,
scale only axis,
tick label style={font=\scriptsize},
     y tick label style={/pgf/number format/.cd,%
          scaled y ticks = false,
          set thousands separator={},
          fixed},
xlabel={\scriptsize Link distance $d$ [km]},
ylabel={\scriptsize Probabilty},
xtick={1000,3000},
ylabel style={at={(2.8ex,5ex)}},
xlabel style={at={(0.5,1.5ex)}},
xticklabels={{1000},{3000}},
xmajorgrids,
ymajorgrids,
xmin=250,
xmax=3700,
ymin=0,
ymax=0.15,
legend columns=1,
legend style={at={(axis cs:1470,0.127)},anchor=west,nodes={scale=0.4, transform shape}}
]
 
  \addlegendimage{Set1-B}
 \addlegendentry{NSFNET};
  \addlegendimage{Set1-A}
 \addlegendentry{ACMN};

 \addplot [ybar interval, fill=Set1-B, fill opacity=0.2, draw=Set1-B, area legend]
  table[]{figures/dist-1.tsv};
  \addplot [Set1-A, thick]
  table[]{figures/dist-2.tsv};

\end{axis}
\end{tikzpicture}
\caption{Flow chart of ACMN generation and throughput estimation. 1000 distinct topologies were generated where each was assigned 100 different link distance realisations. The throughput relaxation $x\in \left[0,1\right]$ was swept to maximise total network throughput and to trade-off individual lightpath capacity for decreased network congestion (i.e. increased $N_\lambda$).}
\label{fig:networkscheme}
\end{figure}
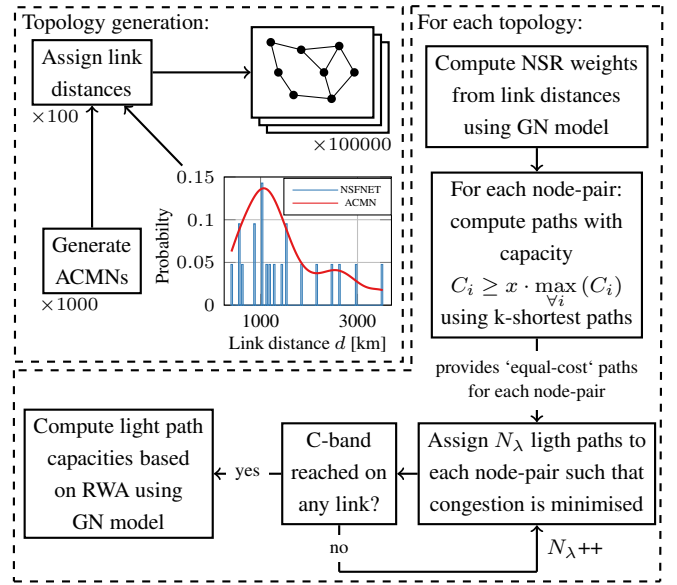

\section{Methodology and throughput estimation}
To study the relationship between topology and network throughput, 1000 distinct logical topologies were generated with 14 nodes and 21 links each, as variants of the NSFNET topology \cite{NSFNET}, chosen as the reference network. Physical link lengths  were randomly assigned to each network link (edge) according to a probabilty density function (PDF), as shown in Fig. 1. The link distances of NSFNET in \cite{Ives} were converted to a continuous PDF by means of kernel density estimation. Using this approach, the generated ACMNs had similar number of nodes, links as well as physical footprint to the NSFNET topology. Each \textit{logical} topology was assigned 100 different distance realisations.
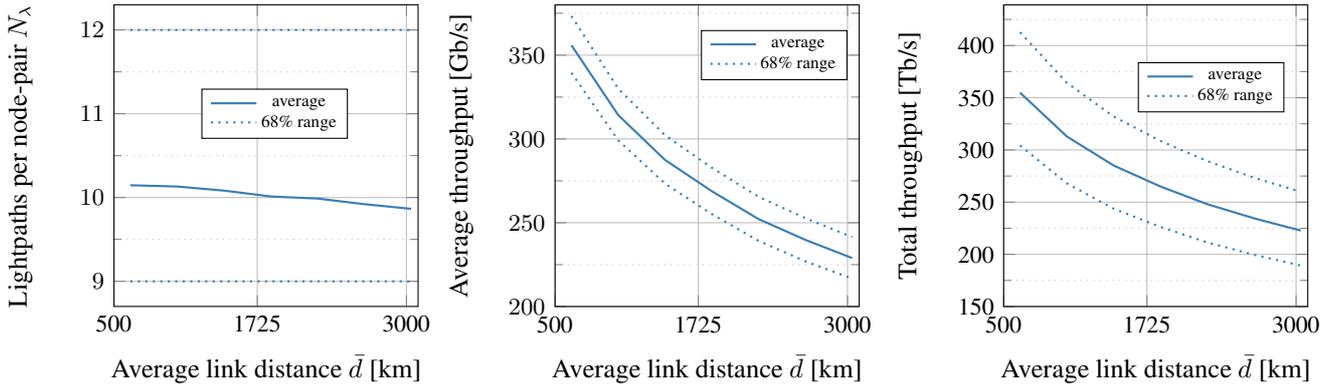
\begin{figure*}[h]
\centering
\begin{tikzpicture}
\pgfplotsset{every tick label/.append style={font=\small}},
\begin{axis}[%
width=4cm,
height=4cm,
at={(0cm,0cm)},
scale only axis,
xlabel={Average link distance $\bar{d}$ [km]},
ylabel={Lightpaths per node-pair $N_\lambda$},
xtick={500, 1725, 3000},
xticklabels={{500}, {1725}, {3000}},
xmajorgrids,
xminorgrids,
ymajorgrids,
yminorgrids,
xminorticks=true,
yminorticks=true,
minor x tick num=1,
minor y tick num=1,
minor grid style={dotted},
xmin=500,
xmax=3100,
legend columns=1,
legend style={at={(axis cs:2500,11)},anchor=east,nodes={scale=0.65, transform shape}}
]

  \addlegendimage{Set1-B, thick}
 \addlegendentry{average};
   \addlegendimage{Set1-B, dotted,thick}
 \addlegendentry{68\% range};

  \addplot [Set1-B, dotted, thick]
  table[]{figures/scale_w-1.tsv};
    \addplot [Set1-B, dotted, thick]
  table[]{figures/scale_w-2.tsv};
    \addplot [Set1-B, thick]
  table[]{figures/scale_w-3.tsv};

\end{axis}
\begin{axis}[%
width=4cm,
height=4cm,
at={(5.8cm,0cm)},
scale only axis,
xlabel={Average link distance $\bar{d}$ [km]},
ylabel={Average throughput [Gb/s]},
xtick={500, 1725, 3000},
xticklabels={{500}, {1725}, {3000}},
ytick={150, 200, 250, 300, 350, 400},
xmajorgrids,
xminorgrids,
ymajorgrids,
yminorgrids,
xminorticks=true,
yminorticks=true,
minor x tick num=1,
minor y tick num=1,
minor grid style={dotted},
xmin=500,
xmax=3100,
ymin=200,
ymax=380,
legend columns=1,
legend style={at={(axis cs:3000,350)},anchor=east,nodes={scale=0.65, transform shape}}
]

  \addlegendimage{Set1-B, thick}
 \addlegendentry{average};
   \addlegendimage{Set1-B, dotted,thick}
 \addlegendentry{68\% range};

  \addplot [Set1-B, thick]
  table[]{figures/scale_t-1.tsv};
    \addplot [Set1-B, dotted, thick]
  table[]{figures/scale_t-2.tsv};
    \addplot [Set1-B, dotted, thick]
  table[]{figures/scale_t-3.tsv};

\end{axis}
\begin{axis}[%
width=4cm,
height=4cm,
at={(11.7cm,0cm)},
scale only axis,
xlabel={Average link distance $\bar{d}$ [km]},
ylabel={Total throughput [Tb/s]},
xtick={500, 1725, 3000},
xticklabels={{500}, {1725}, {3000}},
ytick={150000, 200000, 250000, 300000, 350000, 400000},
yticklabels={{150},{200},{250},{300},{350},{400}},
xmajorgrids,
xminorgrids,
ymajorgrids,
yminorgrids,
xminorticks=true,
yminorticks=true,
minor x tick num=1,
minor y tick num=1,
minor grid style={dotted},
scaled y ticks=false,
xmin=500,
xmax=3100,
ymin=150000,
legend columns=1,
legend style={at={(axis cs:2900,360000)},anchor=east,nodes={scale=0.65, transform shape}}
]

  \addlegendimage{Set1-B, thick}
 \addlegendentry{average};
   \addlegendimage{Set1-B, dotted,thick}
 \addlegendentry{68\% range};

  \addplot [Set1-B, thick]
  table[]{figures/scale_sumt-1.tsv};
    \addplot [Set1-B, dotted, thick]
  table[]{figures/scale_sumt-2.tsv};
    \addplot [Set1-B, dotted, thick]
  table[]{figures/scale_sumt-3.tsv};

\end{axis}
\end{tikzpicture}%
\caption{Established lightpaths per node-pair, their average throughput and total network throughput as a function of average network link distance $\bar{d}$.}
\label{fig:scale}
\end{figure*}
\par 
\ 
All links were assumed to be multiple of 80~km standard single mode fibre spans with parameters $\alpha=0.2\frac{\text{dB}}{\text{km}}$, $D=18\frac{\text{ps}}{\text{nm}\cdot \text{km}}$ and $\gamma=1.2\frac{\text{1}}{\text{W}\cdot\text{km}}$. For this work, the optical bandwidth was constrained to C-band (5~THz). Colourless, directionless and contentionless reconfigurable optical add-drop multiplexers (ROADM) and Erbium-doped fibre amplifiers with 4 dB noise figure were assumed. $32$~GBd, Nyquist-spaced WDM channels were considered and filtering effects and insertion loss of ROADMs were neglected.
\par 
\ 
For each \textit{physical} topology, the throughput was calculated following  in Fig. 1. The overall objective was to maximise the total network throughput. It was assumed that all 91 node-pairs are assigned the same number of lightpaths $N_\lambda$, which is then maximised within C-band. First, a noise-to-signal ratio (NSR), defined as inverse signal-to-noise ratio $\text{NSR}=\frac{1}{\text{SNR}}$, was assigned to each network link. The advantage of working with NSR is that, the total NSR of a lightpath is simply the sum of its link NSRs. This allows the computation of the k-highest SNR paths for a given node-pair using a weighted k-shortest path algorithm. The NSR was calculated using a closed-form approximation of the Gaussian Noise (GN) model \cite{cfJLT}, modelling nonlinear interference as additive white Gaussian noise (AWGN) and capable of accounting for variably loaded network links. For the initial link NSR assignment, however, uniform signal spectra were assumed. The launch power was optimised for every channel in each link following the LOGON strategy \cite{LOGON}.
\par 
\ 
Based on the link NSR, lightpath capacities $C_i$ were estimated using Shannon's formula \cite{shannon}, yielding the maximised lightpath throughput for a given SNR. For each node-pair, NSR weighted k-shortest paths were used to determine all possible lightpaths within a certain capacity range. In particular, only lightpaths that satisfied $C_i\geq x\cdot\underset{\forall i}{\max}\left(C_i\right)$ were kept for a throughput relaxation $x\in\left[0,1\right]$ and where $i$ ranges over \textit{all} possible lightpaths for a given node-pair. Increasing throughput relaxation $x$ generates a larger optimisation space, i.e more 'equal-cost' paths, for the routing and wavelength assignment (RWA) stage. For larger $x$, lightpaths can be rerouted over longer routes leading to lower network congestion and to more established lightpaths across C-band. However, longer routes exhibit lower throughput per lightpath. For this reason, $x$ was swept and optimised for each physical topology to maximise  total network throughput.
\par 
\ 
The set of 'equal-cost' paths for each node-pair is then passed to the RWA. The objective of the RWA is to find a lightpath configuration that minimises the network congestion and in turn maximises the number of established lightpaths per node-pair $N_\lambda$ within C-band. This is done by rerouting lightpaths over longer paths such that physical wavelengths can be reused over physically diverse paths, without collision. A heuristic, which maximises $N_\lambda$, within a constrained optical bandwidth, was proposed in \cite{Baroni}. For each node pair, an arbitrary set of 'equal-cost' paths is required and the heuristic returns a wavelength and route configuration which minimises spectral occupation across all network links for a given $N_\lambda$. The heuristic has been shown to match results of ILP formulations for \textit{logical} topologies, without taking physical transmission properties into account. Here, the heuristic is extended to include physical properties by carefully selecting the set of 'equal-cost' paths according to their physical transmission capacities. After each iteration of the RWA, $N_\lambda$ is incremented until the entire C-band is filled. In summary, the proposed approach finds the maximum number of lightpaths for each node-pair $N_\lambda$ that can be assigned over C-band, for a given optimisation space of 'equal-cost' paths, determined by the throughput relaxation $x$.
\par 
\ 
Finally, the throughput of each lightpath is computed based on the exact lightpath configuration returned by the RWA algorithm. The throughput of a particular lightpath is, hence, dependent on the transmitted distance, fibre parameters as well as number and channel spacing of co-propagating lightpaths in each link.
\begin{figure*}[h]
\centering
\begin{tikzpicture}
\pgfplotsset{every tick label/.append style={font=\small}},
\begin{axis}[%
width=4cm,
height=4cm,
at={(0cm,0cm)},
scale only axis,
xlabel={Normalized network diameter $D$},
ylabel={Lightpaths per node-pair $N_\lambda$},
xtick={1.7, 2.3, 3},
xmajorgrids,
xminorgrids,
ymajorgrids,
yminorgrids,
xminorticks=true,
yminorticks=true,
minor x tick num=1,
minor y tick num=1,
minor grid style={dotted},
xmin=1.6,
xmax=3.1,
legend columns=1,
legend style={at={(axis cs:2.5,11)},anchor=east,nodes={scale=0.65, transform shape}}
]

  \addlegendimage{Set1-A, thick}
 \addlegendentry{average};
   \addlegendimage{Set1-A, dotted,thick}
 \addlegendentry{68\% range};

  \addplot [Set1-A, dotted, thick]
  table[]{figures/banana_w-1.tsv};
    \addplot [Set1-A, dotted, thick]
  table[]{figures/banana_w-2.tsv};
    \addplot [Set1-A,  thick]
  table[]{figures/banana_w-3.tsv};

\end{axis}
\begin{axis}[%
width=4cm,
height=4cm,
at={(5.8cm,0cm)},
scale only axis,
xlabel={Normalized network diameter $D$},
ylabel={Average throughput [Gb/s]},
xtick={1.7, 2.3, 3},
xmajorgrids,
xminorgrids,
ymajorgrids,
yminorgrids,
xminorticks=true,
yminorticks=true,
minor x tick num=1,
minor y tick num=1,
minor grid style={dotted},
xmin=1.6,
xmax=3.1,
ymin=270,
ymax=310,
legend columns=1,
legend style={at={(axis cs:2.5,305)},anchor=east,nodes={scale=0.65, transform shape}}
]

  \addlegendimage{Set1-A, thick}
 \addlegendentry{average};
   \addlegendimage{Set1-A, dotted,thick}
 \addlegendentry{68\% range};

  \addplot [Set1-A, thick]
  table[]{figures/banana_t-1.tsv};
    \addplot [Set1-A, dotted, thick]
  table[]{figures/banana_t-2.tsv};
    \addplot [Set1-A, dotted, thick]
  table[]{figures/banana_t-3.tsv};

\end{axis}
\begin{axis}[%
width=4cm,
height=4cm,
at={(11.7cm,0cm)},
scale only axis,
xlabel={Normalized network diameter $D$},
ylabel={Total throughput [Tb/s]},
xtick={1.7, 2.3, 3},
ytick={240000, 260000, 280000, 300000, 320000, 340000},
yticklabels={{240},{260},{280},{300},{320},{340}},
xmajorgrids,
xminorgrids,
ymajorgrids,
yminorgrids,
xminorticks=true,
yminorticks=true,
minor x tick num=1,
minor y tick num=1,
minor grid style={dotted},
scaled y ticks=false,
xmin=1.6,
xmax=3.1,
legend columns=1,
legend style={at={(axis cs:2.5,300000)},anchor=east,nodes={scale=0.65, transform shape}}
]

  \addlegendimage{Set1-A, thick}
 \addlegendentry{average};
   \addlegendimage{Set1-A, dotted,thick}
 \addlegendentry{68\% range};

  \addplot [Set1-A, thick]
  table[]{figures/banana_sumt-1.tsv};
    \addplot [Set1-A, dotted, thick]
  table[]{figures/banana_sumt-2.tsv};
    \addplot [Set1-A, dotted, thick]
  table[]{figures/banana_sumt-3.tsv};

\end{axis}
\end{tikzpicture}%
\caption{Established lightpaths per node-pair, their average throughput and the total network throughput as a function of normalised network diameter $D$. The normalised network diameter is a measure for the ellipticity of a physical network topology.}
\label{fig:banana}
\end{figure*}
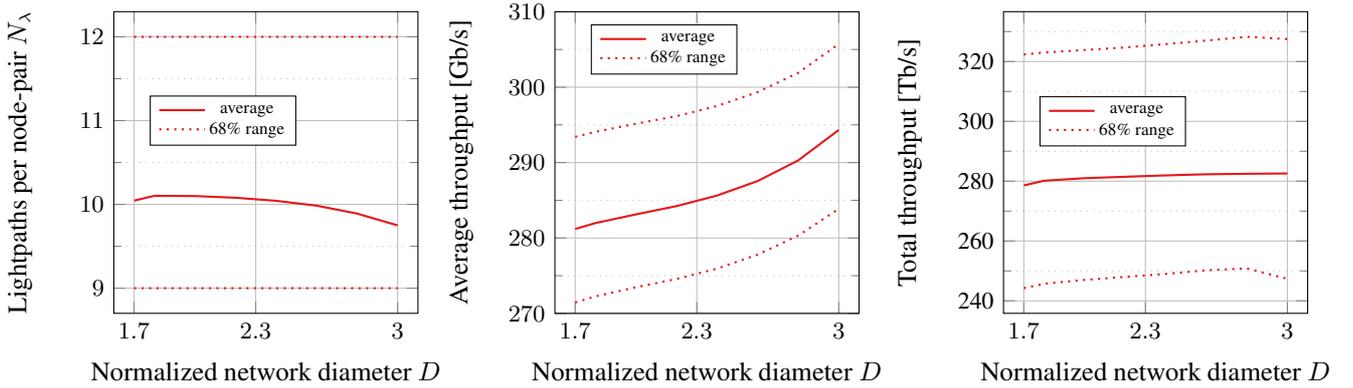
\section{Scaling the size of optical networks}
To study the impact of physical size on network throughput, the average link distance of each ACMN was scaled from 640~km to 3040~km, by scaling the assigned link distances by the factor $\frac{\bar{d}}{1463\text{km}}$, where $\bar{d}$ is the average link distance. The used average link distance of NSFNET is 1463~km. For comparison the average link distances, estimated as in \cite{Ives}, of the DTAG/T-Systems \cite{dtag} and the Google B4 \cite{google} topology are 236~km and 3910~km, respectively.
\par 
\ 
The assigned lightpaths per node pair $N_\lambda$, the average throughput of all lightpaths and the total network throughput are shown Fig. \ref{fig:scale}. As 100,000 \textit{physical} topologies were studied, only the average and the range of 68\% of all realisations are shown. Transmission in mesh optical networks is different to point-to-point transmission, due to variably loaded links. However, the throughput $C_i$ in a single point-to-point link decreases approximately logarithmically with the number of identical spans $n_s$ as $C_i \propto \log_2\left(1+\text{SNR}_{n_s}\right) \approx  \log_2\left(\text{SNR}_1\right) - \log_2\left({n_s}\right)$. Fig \ref{fig:scale}a) shows that, as the physical network size increases, the congestion increases and, therefore, the number of lightpaths per node pair $N_\lambda$ is reduced. This is because as the overall  link distance increases, the number of 'equal-cost' paths is reduced, due to the logarithmic throughput scaling. Thus the RWA heuristic must optimise over fewer 'equal-cost' paths, resulting in worse load balancing and fewer  lightpaths. The average throughput of all lightpaths established decreases (approximately) logarithmically with average link distance, as in Fig. \ref{fig:scale}b), similar to the scaling in a point-to-point transmission. As $N_\lambda$ decreases only marginally with average link distance, in terms of the average throughput, Fig. \ref{fig:scale}c), too, shows an approximate logarithmic decrease of total network throughput with physical network size.
\par 
\ 
Based on this analysis, it can be concluded that the precise routing  only marginally depends on the (uniform) physical network size, but the average throughput per lightpath, as well as the total throughput, decrease logarithmically with the physical network footprint.
\section{Scaling the ellipticity in optical networks}
Another important property of a physical network is its shape. In this work we denote it by the term 'ellipticity' and define a normalised network diameter as $D=\frac{\text{network diameter }\left[\text{km}\right]}{\text{average node-pair distance}\left[\text{km}\right]}$, where the network diameter is the maximum shortest path distance in km over all node-pairs. If the diameter is then normalised by the average node-pair distance,  it can serve as way of parameterising the impact of network shape and defines the ellipticity of a topology. In this work, for each logical topology, link distances were randomly assigned and only kept if they satisfied a fixed normalised network diameter $D$. The average node-pair distance was fixed to 3070~km, which is that of the used NSFNET topology. For comparison, NSFNET has $D=2.13$, the DTAG/T-systems topology and the Google B4 topologies have $D=2.04$ and $D=2.79$. 
\par 
\ 
Fig. \ref{fig:banana} shows the results for the number of established lightpaths per node pair, average lightpath throughput and total network throughput as a function of ellipticity. It can be seen that Fig. \ref{fig:banana}a) shows a marginal decrease in the number of the established lightpaths, thus increasing congestion, as the ellipticity increases. Fig. \ref{fig:banana}b) shows an increase of the average lightpath throughput. Increasing ellipticity increases the distances for some lightpaths, and decreases the distances of the remaining ones. To aid the interpretation of the results, we looked at the scaling of the average throughput for two point-to-point links, where one is increased and the other decreased by $\Delta n_s$ fibre spans. The average throughput of two point-to-point links scales as $\bar{C}_i \appropto 2\log_2\left(\text{SNR}_1\right) - \log_2\left(n_s^2-\Delta n_s^2\right)$ which, indeed, predicts an increase in average lightpath throughput as $\Delta n$ (the ellipticity) is increased. The marginally decreased $N_\lambda$ is balancing the effect of increased average throughput, as ellipticity grows. Thus, the total throughput  only marginally depends on ellipticity, as shown in Fig. \ref{fig:banana}c). 
\ 
\par 
The analysis shows that throughout in optical networks is only weakly dependent on the topology ellipticity, for the ranges studied in this paper.
\section{Conclusion}
The relationship between physical network topology and throughput was studied by parametrising and analysing arbitrarily-connected mesh optical networks. It was found that, similarly to point-to-point links, network throughput decreases approximately logarithmically with physical network size while there is only a minor throughput dependence on the ellipticity of a topology. The study further suggests that network congestion only marginally increases with a uniform increase of physical network size and ellipticity.
%

\vspace*{6pt}
\textit{\footnotesize{
   Support for this work is from UK EPSRC under DTG PhD studentship to D.~Semrau and EPSRC TRANSNET Programme Grant.
}}


\clearpage 

\section*{References}

\vspace*{6pt}
\begin{thebibliography}{17}
\setcounter{NAT@ctr}{0}
\bibitem{nw1}
Ramamurthy, B., Datta, D., Feng, H., et al.: "Impact of transmission impairments on the teletraffic performance of wavelength-routed optical networks," Journal of Lightwave Technology, 1999, 17, (10), pp. 1713--1723
\bibitem{nw2}
Velasco, L., Klinkowski, M., Ruiz, M., et al.: "Modeling the routing and spectrum allocation problem for flexgrid optical networks," Photon. Netw. Commun., 2012, 24, (3), pp. 177--186
\bibitem{nw7}
Sartzetakis, I., Christodoulopoulos, K., Tsekrekos , C. P., et al.: "Quality of transmission estimation in WDM and elastic optical networks accounting for space–spectrum dependencies," Journal of Optical Communications and Networking, 2016, 8, (9), pp. 676--688
\bibitem{nw4}
Zhao, J., Wymeersch, H., Agrell, E.: "Nonlinear impairment-aware static resource allocation in elastic optical networks," Journal of Lightwave Technology, 2015, 33, (22), pp. 4554--4564
\bibitem{nw3}
Roberts, I., Kahn, J. M.: "Efficient discrete rate assignment and power optimization in optical communication systems following the Gaussian noise model," Journal of Lightwave Technology, 2017, 35, (20), pp. 4425--4437
\bibitem{Ives}
Ives, D., Bayvel, P., Savory, S.: "Adapting transmitter power and modulation format to improve optical network performance utilizing the Gaussian noise model of nonlinear impairments," Journal of Lightwave Technology, 2014, 32, (21), pp. 3485--3494
\bibitem{nw5}
Tessinari, R. S., Paiva, M. H. M., Monteiro, M. E. et al.:"On the impact of the physical topology on the optical network performance," IEEE British and Irish Conference on Optics and Photonics, London, December, 2018
\bibitem{Baroni}
Baroni, S., Bayvel, P.: "Wavelength requirements in arbitrarily connected wavelength-routed optical networks," Journal of Lightwave Technology, 1997, 15, (2), pp. 242--251
\bibitem{NSFNET}
Ramaswami, R., Sivarajan, K. N.: "Design of logical topologies for wavelength-routed all-optical network," Proc. Infocom, 1995, pp. 1316--1325 
\bibitem{cfJLT}
Semrau, D., Killey, R. I., Bayvel, P.: "A closed-form approximation of the Gaussian noise model in the presence of inter-channel stimulated Raman scattering," Journal of Lightwave Technology, 2019, 37, (9), pp. 1924--1936
\bibitem{LOGON}
Poggiolini, P., Bosco, G., Carena, A., et al.:"The LOGON strategy for low-complexity control plane implementation in new-generation flexible networks," Proc. Optical Fiber Communication Conference, Anaheim, March, 2013, OW1H.3, pp. 1--3
\bibitem{shannon}
Shannon, C. E.: "A mathematical theory of communication," Bell System Technical Journal, 1948, 27, pp. 379--423
\bibitem{dtag}
Monoyios, D., Vlachos, K.: "A multiobjective genetic algorithms for solving the impairment-aware routing and wavelength assignment problem," Journal of Optical Communications and Networking, 2011, 3, (1), pp. 40--47
\bibitem{google}
Jain, S., Zhu, M., Zolla, J., et al.: "B4: Experience with a globally-deployed software defined WAN," Journal of Optical Communications and Networking", Special Interest Group on Data Communication, Hong Kong, August, 2013, pp. 3--14
\end{thebibliography}
\end{document}